\documentclass{SciPost}

\hypersetup{
    colorlinks,
    linkcolor={red!50!black},
    citecolor={blue!50!black},
    urlcolor={blue!80!black}
}

\usepackage[normalem]{ulem}
\usepackage[bitstream-charter]{mathdesign}
\usepackage{subcaption}
\usepackage{booktabs}

\DeclareSymbolFont{usualmathcal}{OMS}{cmsy}{m}{n}
\DeclareSymbolFontAlphabet{\mathcal}{usualmathcal}

\fancypagestyle{SPstyle}{
\fancyhf{}
\lhead{\colorbox{scipostblue}{\bf \color{white} ~SciPost Physics Community Reports }}
\rhead{{\bf \color{scipostdeepblue} ~Submission }}

\fancyfoot[C]{\textbf{\thepage}}
}

\def\muR {\mu_R}
\def\muF {\mu_F}
\def\mH{{m_H^{}}}
\def\muRs{{\mu_R^{\,2}}}
\def\muFs{{\mu_F^{\,2}}}
\def\mHs{{m_H^{\,2}}}
\def\MZ{{m_Z}}

\def\mt{{m_t}}

\def\muR {\mu_R}
\def\muF {\mu_F}
\def\mH{{m_H^{}}}
\def\muRs{{\mu_R^{\,2}}}
\def\muFs{{\mu_F^{\,2}}}
\def\mHs{{m_H^{\,2}}}

\newcommand{\nn}{\nonumber}

\newcommand{\as}{\alpha_{\rm s}}

\def\zt{{(1\!-\!z)}}

\usepackage{array}
\newcolumntype{P}[1]{>{\centering\arraybackslash}p{#1}}

\usepackage[normalem]{ulem}

\newcommand{\sect}[1]{section\ \ref{#1}}

\begin{document}

\pagestyle{SPstyle}
%%%
%{\flushleft{P3H-25-080, TTK-25-34}}
%%%
\begin{center}{\Large \textbf{\color{scipostdeepblue}{
%%%%%%%%%% TODO: Write your article's title here
The inclusive Higgs boson cross-section in gluon-gluon fusion \\[1ex] 
in soft-virtual approximation at fourth order in QCD
%%%%%%%%%% END TODO: TITLE
}}}\end{center}

\begin{center}
\textbf{
%%%%%%%%%% TODO: AUTHORS
% Write the author list here. 
% Use (full) first name (+ middle name initials) + surname format.
% Separate subsequent authors by a comma, omit comma and use "and" for the last author.
% Mark the corresponding author(s) with a superscript symbol in this order
% \star, \dagger, \ddagger, \circ, \S, \P, \parallel, ...
Goutam Das\textsuperscript{1$\star$} and 
Sven-Olaf Moch\textsuperscript{2$\dagger$}
%%%%%%%%%% END TODO: AUTHORS
}
\end{center}

\begin{center}
%%%%%%%%%% TODO: AFFILIATIONS
% Write all affiliations here.
% Format: institute, city, country
{\bf 1} Institut für Theoretische Teilchenphysik und Kosmologie, RWTH Aachen University,  D-52056 Aachen, Germany\\
{\bf 2} II. Institut für Theoretische Physik, Universit\"at Hamburg, Luruper Chaussee 149,\\
D-22761 Hamburg, Germany
%%%%%%%%%% END TODO: AFFILIATIONS
%%%%%%%%%% TODO: EMAIL
% Provide email address of corresponding author(s)
\\[\baselineskip]
$\star$~\href{mailto:goutam@physik.rwth-aachen.de}{\small goutam@physik.rwth-aachen.de}\,,\quad
$\dagger$~\href{mailto:sven-olaf.moch@desy.de}{\small sven-olaf.moch@desy.de}\,\quad
%%%%%%%%%% END TODO: EMAIL
\end{center}

\section*{\color{scipostdeepblue}{Abstract}}
\textbf{\boldmath{%
%%%%%%%%%% TODO: ABSTRACT
% Write your abstract here.
    We present precise results for the inclusive Higgs boson cross-section in gluon-gluon fusion at the LHC 
    considering state-of-the-art fourth-order results in perturbative QCD
    arising from the dominant soft and virtual gluon emissions.
    Utilizing four-loop QCD results for the gluon-form factor, the splitting function and related anomalous dimensions,
    we study the effects of threshold enhanced soft gluon emissions and estimate their impact on the total cross-section at the fourth order.
    Our study highlights the role of these higher-order contributions in improving the perturbative convergence and in significantly reducing the renormalization and factorization scale uncertainties.
    The results provide strong evidence for the perturbative stability and reliability of Higgs boson cross-section predictions at the LHC, thereby reinforcing the robustness of theoretical inputs in precision Higgs phenomenology.
    We also provide cross-section predictions using a large set of available
    parton distribution functions and show that, together with the value of
    the strong coupling $\as(m_Z)$, they cause the largest residual uncertainty
    for the Higgs boson cross-section in gluon-gluon fusion.
%%%%%%%%%% END TODO: ABSTRACT
}
}

\vspace{\baselineskip}

%%%%%%%%%% BLOCK: Copyright information
% This block will be filled during the proof stage, and finilized just before publication.
% It exists here only as a placeholder, and should not be modified by authors.
\noindent\textcolor{white!90!black}{%
\fbox{\parbox{0.975\linewidth}{%
\textcolor{white!40!black}{\begin{tabular}{lr}%
  \begin{minipage}{0.6\textwidth}%
    {\small Copyright attribution to authors. \newline
    This work is a submission to SciPost Phys. Comm. Rep. \newline
    License information to appear upon publication. \newline
    Publication information to appear upon publication.}
  \end{minipage} & \begin{minipage}{0.4\textwidth}
    {\small Received Date \newline Accepted Date \newline Published Date}%
  \end{minipage}
\end{tabular}}
}}
}
%%%%%%%%%% BLOCK: Copyright information

%%%%%%%%%% TODO: LINENO
% For convenience during refereeing we turn on line numbers:
%\linenumbers
% You should run LaTeX twice in order for the line numbers to appear.
%%%%%%%%%% END TODO: LINENO

%%%%%%%%%% TODO: TOC 
% Guideline: if your paper is longer that 6 pages, include a TOC
% To remove the TOC, simply cut the following block
\vspace{10pt}
\noindent\rule{\textwidth}{1pt}
\tableofcontents
\noindent\rule{\textwidth}{1pt}
\vspace{10pt}
%%%%%%%%%% END TODO: TOC

%%%%%%%%% TODO: CONTENTS 

\section{Introduction}
%%% WHY Higgs?
The discovery of the Higgs boson at the Large Hadron Collider (LHC) has confirmed a central pillar of the Standard Model and simultaneously opened new avenues for exploring physics beyond it.  Understanding the properties of the Higgs boson is therefore essential in the search for potential signs of new physics in both current and future LHC programs. Precise theoretical predictions, combined with high-precision experimental measurements, are key to enhancing our sensitivity to possible new phenomena that may be coupled to the Higgs sector.

%%% WHY 4th order?
At the LHC, the Higgs boson is dominantly produced through the gluon-gluon
fusion and receives substantial perturbative corrections in quantum
chromodynamics (QCD)~\cite{Dawson:1990zj,Djouadi:1991tka,Spira:1995rr,Harlander:2005rq,
  Harlander:2002wh,Anastasiou:2002yz,Ravindran:2003um,
  Anastasiou:2015vya,Mistlberger:2018etf}.
Indeed, the perturbative series for the Higgs boson production cross-section is known to converge slowly at lower orders.
For instance, the next-to-leading order (NLO) QCD correction nearly doubles the leading order (LO) result. The next-to-next-to-leading order (NNLO) adds
an additional $\sim 25\%$ relative to NLO, while the next-to-next-to-next-to-leading order (N3LO) contributes a further $\sim 3.5\%$ on top of the NNLO result.
At N3LO,  the residual theoretical uncertainty due to variations of the
renormalization and factorization scales $\muR$ and $\muF$
-- estimated using their standard seven-point variation -- is approximately $4\%$. 
To further improve the precision of the Higgs boson cross-section, it is crucial to incorporate fourth-order QCD corrections. These higher-order contributions enhance the perturbative convergence and help reduce the sensitivity to unphysical scale choices, thereby improving the robustness of theoretical predictions.

%%% WHY Threshold?
At the LHC, the Higgs cross-section receives dominant contribution from soft gluon emissions, particularly near threshold where $z =\mHs/\widehat{s} \to 1$, with $\mH$ being the Higgs mass and $\widehat{s}$ being the partonic centre of mass energy.  
The effects of these soft gluons are manifested in terms of well-known large threshold logarithms in $1-z$ which in Mellin-$N$ space, take the form $\ln^k N$ with $1 \geq k \geq 2n$ at order $n$.  
The Mellin-space formulation not only provides a remarkably accurate approximation to the exact fixed-order cross-section at N3LO \cite{Moch:2005ky,deFlorian:2014vta} but also facilitates the resummation of these large logarithms via threshold exponentiation techniques. 
This motivates the construction of an approximate fourth-order cross-section by incorporating the dominant threshold logarithms in Mellin space.

%%% Plan
In this report, we predict the total Higgs boson cross-section in gluon-gluon fusion
at fourth order in QCD using available four-loop QCD results for the gluon form factor, the splitting functions and related anomalous dimensions.
After introducing the basic notation and formulae in \sect{sec:theory}, we
present a comprehensive study of Higgs boson cross-section and its
uncertainties, in \sect{sec:results} relevant for $13.6$ TeV LHC,
also using different sets of parton distribution functions (PDFs), accurate
to NNLO and N3LO.

%\cleardoublepage
%%%%%%%%%%%%%%%%%%%%%%%%%
\section {HEFT cross-section}\label{sec:theory}
%%%%%%%%%%%%%%%%%%%%%%%%%
To compute the fourth order QCD corrections, we use  the Higgs effective
theory (HEFT) and the heavy top-quark mass limit where the Higgs boson couples to gluons via effective interaction given by,
\begin{align} \label{eq:Leff}
	{\cal L}_{\rm eff}  =
	-\frac{1}{4 \upsilon} C(\muR^2)  H G_{\mu\nu}^a G^{\mu\nu}_a
\end{align}
with $\upsilon$ being the Higgs vacuum expectation value, $H$ the Higgs field, and $G_{\mu\nu}^a$ the gluon field strength tensor.
The inclusive Higgs boson production cross-section in the gluon-gluon fusion channel in the heavy top-quark mass limit takes the form,
\begin{eqnarray} \label{eq:had-xsect-master}
	\sigma(S,\mHs) &\!=\!& \tau\: 
	\sum\limits_{a,b} \:\int_0^1 \frac{dx_1^{}}{x_1^{}} \;\frac{dx_2^{}}{x_2^{}} 
	\; f_{a/h_1^{}}(x_1^{},\muFs)\;f_{b/h_2^{}}(x_2^{},\muFs) 
	\int_0^1 \! dz \;\delta \Big(z - \frac{\tau}{x_1^{} x_2^{}}\Big)\, \times\,  
	\nn \\[0.5mm] & & 
	\times\:\: \widetilde{\sigma}_0^{H} \: c^H_{ab}(z,\, \as(\muRs),\, 
	\mHs/\muRs,\, \mHs/\muFs) \;\; ,
\end{eqnarray}
where the Born factor in the large $\mt \to \infty$ limit reads
\begin{eqnarray}
	\label{eq:sigma0}
	\widetilde{\sigma}_0^{H} \:\: = \:\:
	\frac{\pi\, C(\muRs)^2}{8\, n^{ }_{\!A}\, \upsilon^2} \quad \mbox{ with } \quad
	C(\muRs) \:\: = \:\: - \,\frac{\as(\muRs)}{3 \pi} 
	\: \Big\{ 1 \,+\, 11\: \frac{\as(\muRs)}{4 \pi} \: + \: \ldots \Big\} \;\; .
\end{eqnarray}
Here $\tau=\mHs/S$, and $f_{a/h}(x,\muFs)$ are the PDFs in the proton.
The coefficient function $c_{ab}^H$ can be expanded in strong coupling $\as$
with the LO term taking the form 
$c_{ab}^{H,(0)} = \delta_{ag}\,\delta_{bg}\,\delta \zt$.

The computation of the coefficient function at fourth order in the threshold limit requires several perturbative ingredients. 
First, the contribution from the gluon form factor, which encodes purely virtual gluon effects, is known~\cite{Agarwal:2021zft,Lee:2022nhh}. 
Second, the fourth-order gluon splitting functions, which describe collinear gluon emissions, are needed. 
These are known to a sufficient degree through the computation of several Mellin moments~\cite{Falcioni:2024qpd}, allowing us to reliably extract the coefficient of the $1/\epsilon$ pole in the soft-virtual approximation.
Furthermore, employing techniques from threshold resummation, we are able to estimate the effects of process-dependent coefficients that appear at this order.\footnote{For a detailed discussion of the fourth-order ingredients and their computation, see \cite{Das:2019btv,Das:2020adl,Kniehl:2025ttz}.} 
In the next section, we compute the cross-section with the contribution of soft and virtual gluons at fourth order.

%\gd{Note, we work in the Born-improved HEFT, i.e. LO is exact and rescaled against the EFT $\mt \to \infty$.
%	$\widetilde{\sigma}_0^H$ should we change? Mentioned in the beginning of next section.}
%%%%%%%%%%%%%%%%%%%%%%%%%%%%%%%
\section{Results}\label{sec:results}
%%%%%%%%%%%%%%%%%%%%%%%%%%%%%%%%

It is well known that the validity of HEFT predictions can be significantly improved by rescaling them with the exact LO result computed using the full theory, which includes the finite top-quark mass and $n_f=5$ massless quark flavors. Accordingly, to enhance the reliability of the QCD predictions incorporating fourth-order soft gluon corrections, we rescale the HEFT results in Eq.~\eqref{eq:had-xsect-master} by the exact LO cross-section.
We present our numerical estimates for proton-proton collisions at $\sqrt{S}=13.6$ TeV at the LHC. The Higgs boson mass is taken to be $\mH=125$ GeV, and the on-shell top-quark mass is set to $\mt=172.5$ GeV. Results up to N3LO are obtained using \texttt{iHixs-2} \cite{Dulat:2018rbf}, whereas N4LOsv results are computed with an in-house code.

For our central scale choices, we use $\muR^c=\muF^c=m_H/2$ , which is known to yield improved perturbative convergence by reducing the size of higher-order corrections. Theoretical uncertainties from scale variations are estimated using the standard seven-point method: we vary the renormalization scale $\muR$
and the factorization scale $\muF$ independently within the range $[\mH/4,\mH]$, subject to the constraint
\begin{align}
	\frac{1}{2} \leq \frac{\muR/\muR^c}{\muF/\muF^c} \leq 2\,,
\end{align}
which leads to the following seven scale combinations (in units of the central scale):

\noindent
$( \muR, \muF) = \{ (1/2,1/2), (1/2,1),(1,1/2),(1,1),(1,2),(2,1),(2,2)\} \times (\muR^c,\muF^c)$.
The final scale uncertainty is then obtained by taking the envelope formed by the maximum and minimum deviations from the central prediction.

The theoretical prediction for the inclusive Higgs boson production cross-section at the LHC, as given in Eq.~\eqref{eq:had-xsect-master}, involves the choice of a particular PDF set.
PDFs, being non-perturbative and determined through fits to a global set of
data, introduce an additional source of theoretical uncertainty, which must be quantified. 
Different PDF collaborations adopt varying methodologies and assumptions in determining these functions, leading to differences in the predicted cross-sections. 
To assess the impact of PDF uncertainties on our results, we consider several
modern NNLO PDF sets accessed via the {\sc LHAPDF} interface~\cite{Buckley:2014ana}:
{\tt ABMP16\_5\_nnlo}~\cite{Alekhin:2017kpj},
{\tt ABMPtt\_5\_nnlo}~\cite{Alekhin:2024bhs}, 
{\tt CT18NNLO}~\cite{Hou:2019efy},
{\tt MSHT20nnlo\_as118}~\cite{Bailey:2020ooq},
{\tt NNPDF40\_nnlo\_as\_01180}~\cite{NNPDF:2021njg},
{\tt PDF4LHC21\_40}~\cite{PDF4LHCWorkingGroup:2022cjn};
and approximate N3LO PDF sets:
\noindent
{\tt MSHT20an3lo\_as118}~\cite{McGowan:2022nag},
{\tt NNPDF40\_an3lo\_as\_01180}~\cite{NNPDF:2024nan}\,.

\noindent
{\bf Summary of the parameters:}
\begin{align}
	\sqrt{S}\equiv E_{\rm CM} &= 13.6 ~\text{ TeV },& \qquad  m_H &= 125 \text{ GeV } ,& \qquad  m_t  &= 172.5& \text{ GeV (on-shell scheme)}
	\nonumber\\
	\muR^c &= m_H/2,& \qquad \muF^c &=  m_H/2,& \qquad n_f &= 5 \,.& 
\end{align}
%\clearpage
\noindent
{\bf Error estimation:\\}
The perturbative correction at order $k$ is quantified by $\Delta({\rm N{\it k}LO})$,
defined as the relative change in the cross-section at order $k$ compared to the previous order $(k-1)$.
\begin{align}
	\Delta({\rm N{\it k}LO}) &= \pm
	\left|\frac{\sigma^{\rm N{\it k}LO}}{\sigma^{\rm N({\it k}-1)LO}} 
	-1 \right| \,.
\end{align}
This quantity serves to illustrate the perturbative convergence and to put the quantitative error estimation into perspective.
In addition to uncertainties arising from scale variations and intrinsic PDF errors, we also account for several other sources of theoretical uncertainty. 
For the $k-$th perturbative order, we define:
\begin{align}
	\delta^{(k)}(\text{Scale}) &=
	\left\{ \max\big(\sigma^{\rm N{\it k}LO}(\muR,\muF) 
	- \sigma^{\rm N{\it k}LO}(\muR^c,\muF^c)\big),
	\min\big(\sigma^{\rm N{\it k}LO}(\muR,\muF) 
	- \sigma^{\rm N{\it k}LO}(\muR^c,\muF^c)\big)
	\right\}\,,
	\nonumber\\
	\delta^{(k)}(\alpha_s) &= 
	\pm\frac{1}{2} 
	\frac{ \big|
		\sigma^{\rm N{\it k}LO}(\alpha^{+}_s(m_Z))
		-\sigma^{\rm N{\it k}LO}(\alpha^{-}_s(m_Z))\big|}{\sigma^{\rm N{\it k}LO}(\alpha^c_s(m_Z))}\,,
	\nonumber\\
	\delta^{(k)}(\text{PDF}+\alpha_s) &= 
	\sqrt{\delta^{(k)}(\text{PDF})^2+\delta^{(k)}(\alpha_s)^2}\,,
	\nonumber\\
	\delta^{(k)}(\text{PDF-TH}) &= 
	\pm\frac{1}{2} \frac{\big|
		\sigma^{\rm N{\it k}LO}(\text{N3LO PDF})
		-\sigma^{\rm N{\it k}LO}(\text{NNLO PDF})\big|}{\sigma^{\rm N{\it k}LO}(\text{N3LO PDF})}\,,
\end{align}
where $\delta^{(k)}(\text{PDF})$ is the intrinsic PDF uncertainty for the perturbative order $k$ and 
$	\delta^{(k)}(\text{Scale}) $ is the seven-point scale uncertainty described earlier. 
The uncertainty due to the strong coupling $\alpha_s$ at order $k$ is estimated by varying 
$\alpha_s(\MZ)$ around its central value $\alpha^c_s(\MZ)$ by $\pm 1\sigma$.
The resulting uncertainty is obtained by taking the absolute difference between the cross-sections computed with
$\alpha^+_s(\MZ)$  and $\alpha^-_s(\MZ)$, and then normalizing this difference to the cross-section obtained using the central value  $\alpha^c_s(\MZ)$. The combined PDF and $\alpha_s(\MZ)$ uncertainties, denoted by
$\delta^{(k)}(\text{PDF}+\alpha_s)$ are obtained by adding the individual PDF and $\alpha_s$ uncertainties in quadrature. To estimate the theoretical uncertainty associated with the choice of PDF sets,  $	\delta^{(k)}(\text{PDF-TH})$  we follow the procedure described above: we compare the cross-section at N3LO using 
PDFs with that using NNLO PDFs, and then divide the difference by two. This estimation is only applicable to the two PDF sets for which N3LO PDFs are available, and we apply it only to the  N4LOsv results, as no N4LO PDFs currently exist. 

\noindent
Below we consider two different scenarios corresponding to choosing $ \alpha^c_s(m_Z)$:
\begin{itemize}
	\item 
	%%%% N3LO
	\begin{table}[ht!]
		\centering{
			\begin{tabular}{|P{3.3cm}|P{0.9cm}|P{1.05cm}|P{1.0cm}|P{0.9cm}|P{0.9cm}|P{1.05cm}|P{1.0cm}|P{0.9cm}|}
				\hline 
				\multicolumn{1}{|c|}{} &
				\multicolumn{4}{c|}{N3LO ($\Delta$, $\delta$ in $\%$)} &
				\multicolumn{4}{c|}{N4LOsv ($\Delta$, $\delta$  in $\%$)} \\
				\hline
				\scalebox{0.8}{ PDF Name }
				& \scalebox{0.8}{ Central }
				& \scalebox{0.8}{ $\hspace*{-2mm} \Delta({\rm N3LO}$)}
				& \scalebox{0.8}{ $\delta(\text{\small Scale})$ }
				& \scalebox{0.8}{ $\delta(\text{\small PDF})$ }
				& \scalebox{0.8}{ Central }
				& \scalebox{0.8}{ $\hspace*{-2mm} \Delta({\rm N4LO}$)}
				& \scalebox{0.8}{ $\delta(\text{\small Scale})$ }
				& \scalebox{0.8}{ $\delta(\text{\small PDF})$ }\\
				\hline
				\hline  \rule{0pt}{2.5ex}
 				{\small \tt ABMP16}\cite{Alekhin:2017kpj} 
				&$48.8$
				&$3.3$
				&$^{+0.2}_{ -3.6}$ 
				&$^{+1.7}_{ -1.7}$
				&$48.7$
				&$-0.1$
				&$^{+0.5}_{ -2.1}$
				&$^{+1.7}_{ -1.7}$\\[1ex]  
				\hline  \rule{0pt}{2.5ex}
				{\small \tt ABMPtt}\cite{Alekhin:2024bhs}
				&$48.4$
				&$3.3$
				&$^{+0.2}_{-3.6}$ 
				&$^{+1.5}_{ -1.5}$
				&$48.4$
				&$-0.1$
				&$^{+0.5}_{ -2.1}$
				&$^{+1.5}_{ -1.5}$\\[1ex]  
				\hline  \rule{0pt}{2.5ex}
				{\small \tt CT18NNLO}\cite{Hou:2019efy}
				&$51.3$
				&$3.5$
				&$^{+0.3}_{-3.9}$ 
				&$^{+2.8}_{ -3.6}$
				&$51.3$
				&$-0.1$
				&$^{+0.5}_{ -2.3}$
				&$^{+2.8}_{ -3.6}$\\[1ex]
				\hline  \rule{0pt}{2.5ex}
				{\small \tt MSHT20}\cite{Bailey:2020ooq}
				&$51.4$
				&$3.5$
				&$^{+0.3}_{-3.9}$ 
				&$^{+1.2}_{ -1.2}$
				&$51.3$
				&$-0.1$
				&$^{+0.5}_{ -2.3}$
				&$^{+1.2}_{ -1.2}$\\[1ex]    
				\hline  \rule{0pt}{2.5ex}
				{\small \tt NNPDF40}\cite{NNPDF:2021njg} 
				&$51.7$
				&$3.5$
				&$^{+0.3}_{-3.9}$  
				&$^{+0.6}_{ -0.6}$
				&$51.7$
				&$-0.1$
				&$^{+0.5}_{ -2.3}$
				&$^{+0.6}_{ -0.6}$\\[1ex]   
				\hline  \rule{0pt}{2.5ex}
				{\small \tt PDF4LHC21}\cite{PDF4LHCWorkingGroup:2022cjn} 
				&$51.6$
				&$3.5$
				&$^{+0.3}_{-3.9}$ 
				&$^{+0.6}_{-0.6}$
				&$51.5$
				&$-0.1$
				&$^{+0.5}_{ -2.3}$
				&$^{+0.6}_{ -0.6}$\\[1ex] 
				\hline  \rule{0pt}{2.5ex}
				{\small \tt MSHT20an3lo}~\cite{McGowan:2022nag}
				&$48.7$
				&$3.5$
				&$^{+0.3}_{-3.9}$ 
				&$^{+1.9}_{ -1.7}$
				&$48.7$
				&$-0.1$
				&$^{+0.5}_{ -2.3}$
				&$^{+1.9}_{ -1.7}$\\[1ex]  
				\hline  \rule{0pt}{2.5ex}
				{\small \tt NNPDF40an3lo}\cite{NNPDF:2024nan}  
				&$50.6$
				&$3.5$
				&$^{+0.3}_{-3.9}$  
				&$^{+0.6}_{-0.6}$
				&$50.6$
				&$-0.1$
				&$^{+0.5}_{ -2.3}$
				&$^{+0.6}_{ -0.6}$\\[1ex]
				\hline
			\end{tabular}
		}
		\caption{ 
			Higgs cross-section at N3LO and N4LOsv for
			 $(\muR^c,\muF^c)=(1/2,1/2)\mH$, $\sqrt{S}=13.6$ TeV, and $\alpha_s^c(\MZ)$ from LHAPDF (NNLO value).}
		\label{tab:inc-xsect1} 
	\end{table}
%%%%%%%%%%%%%%%%%%%%%%%%%%%%	
	{\bf Case-I: $ \alpha^c_s(m_Z)$ from {\sc Lhapdf}:} 
%%%%%%%%%%%%%%%%%%%%%%%%%%%%	
The PDF uncertainty is evaluated using the standard intrinsic method, employing various modern PDF sets accessed via the {\sc LHAPDF} interface. 
It is important to note that the ABMP sets determine the strong coupling
simultaneously with the PDFs, i.e.\
$\alpha_s(m_Z) = 0.1147 \pm 0.0008$ for {\tt ABMP16\_5\_nnlo} 
and
$\alpha_s(m_Z) = 0.1150 \pm 0.0009$ for {\tt ABMPtt\_5\_nnlo}, while all other PDF sets considered here use a fixed value of $\alpha_s(m_Z)=0.118$.
In Table~\ref{tab:inc-xsect1}, we present the results for both N3LO and N4LOsv, including the central values of the cross-sections along with their associated scale and intrinsic PDF uncertainties. The N4LOsv correction to the cross-section is modest, amounting to a reduction of approximately $-0.1\%$ relative to N3LO. However, the associated scale uncertainty is significantly reduced by a factor of $2$ at N4LOsv, while the intrinsic PDF uncertainty remains essentially unchanged. 
Interestingly, the use of approximate N3LO PDFs has a noticeable impact on the central values of the cross-section. For example, the NNPDF-based prediction shifts downward by about $-2\%$ and the MSHT prediction by around $-5\%$,
when compared to the corresponding values obtained using NNLO PDFs.
These shifts reflect the impact of higher order corrections (beyond NNLO) on the
determination of the gluon distribution, but also particular methodologies in
the implementation of QCD theory predictions at N3LO within the respective PDF
fits.
A detailed benchmarking of different treatments of the QCD evolution of
unpolarized PDFs at approximate N3LO has been performed in \cite{Cooper-Sarkar:2024crx}.

%%%%%%%%%%%%%%%%%%%%%%%%%%
	\item {\bf Case-II:  $ \alpha^c_s(m_Z) = 0.118$:} 
%%%%%%%%%%%%%%%%%%%%%%%%%%
The strong coupling is set to the same value of $\alpha^c_s(m_Z) = 0.118$ for all PDF sets.
A $1\sigma$ variation in the strong coupling around its central value leads to
\noindent 
$\alpha^{+}_s(m_Z) \equiv \alpha^{1\sigma+}_s(m_Z) = 0.1195$ and $ \alpha^{-}_s(m_Z) \equiv \alpha^{1\sigma-}_s(m_Z)=0.1165$. 
In Table~\ref{tab:inc-xsect2}, we present the corresponding results at N4LOsv.
Noticeably, the central cross-section values obtained with the ABMP PDF sets shift almost by $10\%$ with the choice $ \alpha^c_s(m_Z) = 0.118$ compared to the values reported in Table~\ref{tab:inc-xsect1}.
This is due the lower nominal value of $\alpha^c_s(m_Z)$ extracted by ABMP in a simultaneous fit with the PDFs.
Across all the PDF sets considered, the dominant contribution to the theoretical uncertainty at this order arises from the $1\sigma$ variation in $\alpha_s$, leading to an estimated uncertainty of approximately $4\%$.
	%%%% N4LO
	%\floatsetup[table]{font=small}
	\begin{table}[ht!]
		\centering{
			\begin{tabular}{|P{3.3cm}|P{1.1cm}|P{1.1cm}|P{1.1cm}|P{1.cm}|P{0.8cm}|P{1.4cm}|P{1.2cm}|}
				\hline 
				\multicolumn{1}{|c|}{$k=4$} &
				\multicolumn{7}{c|}{N4LOsv ($\Delta$, $\delta$ in $\%$)} \\
				\hline
				\scalebox{0.8}{PDF Name}
				& \scalebox{0.8}{Central}
				& \scalebox{0.8}{$\hspace*{-1mm} \Delta({\rm N4LO}$)}
				& \scalebox{0.8}{$\delta(\text{\small Scale})$}
				& \scalebox{0.8}{$\delta(\text{\small PDF})$}
				& \scalebox{0.8}{$\delta(\alpha_s)$}
				& \scalebox{0.8}{$\delta(\text{\small PDF}+\alpha_s)$} 
				& \scalebox{0.8}{$\delta(\text{\small PDF-TH})$} \\
				\hline
				\hline  \rule{0pt}{2.5ex}
				{\small \tt ABMP16}\cite{Alekhin:2017kpj} 
				&$53.2$
				&$-0.1$
				&$^{+0.4}_{-2.3}$ 
				&$^{+1.0}_{-1.0}$
				&$\pm 4.0$
				&$^{+4.1}_{-4.1}$  
				&$-$\\[1ex]  
				\hline  \rule{0pt}{2.5ex}
				{\small \tt ABMPtt}\cite{Alekhin:2024bhs}
				&$52.4$
				&$-0.1$
				&$^{+0.4}_{-2.3}$ 
				&$^{+1.1}_{-1.1}$
				&$\pm 4.0$
				&$^{+4.1}_{-4.1}$ 
				&$-$\\[1ex]  
				\hline  \rule{0pt}{2.5ex}
				{\small \tt CT18NNLO}\cite{Hou:2019efy}
				&$51.3$
				&$-0.1$
				&$^{+0.5}_{-2.3}$ 
				&$^{+2.8}_{-3.6}$
				&$\pm 4.0$
				&$^{+4.9}_{-5.4}$
				&$-$\\[1ex]  
				\hline  \rule{0pt}{2.5ex}
				{\small \tt MSHT20nnlo}\cite{Bailey:2020ooq}
				&$51.3$
				&$-0.1$
				&$^{+0.5}_{-2.3}$ 
				&$^{+1.2}_{-1.2}$
				&$\pm 4.0$
				&$^{+4.2}_{-4.2}$
				&$-$\\[1ex]  
				\hline  \rule{0pt}{2.5ex}
				{\small \tt NNPDF40nnlo}\cite{NNPDF:2021njg} 
				&$51.7$
				&$-0.1$
				&$^{+0.5}_{-2.3}$ 
				&$^{+0.6}_{-0.6}$
				&$\pm 4.0$
				&$^{+4.0}_{-4.0}$
				&$-$\\[1ex]  
				\hline  \rule{0pt}{2.5ex}
				{\small \tt PDF4LHC21}\cite{PDF4LHCWorkingGroup:2022cjn} 
				&$51.5$
				&$-0.1$
				&$^{+0.5}_{-2.3}$ 
				&$^{+1.7}_{-1.7}$
				&$\pm 4.0$
				&$^{+4.3}_{-4.3}$
				&$-$\\[1ex]  
				\hline  \rule{0pt}{2.5ex}
				{\small \tt MSHT20an3lo}~\cite{McGowan:2022nag}
				&$48.7$
				&$-0.1$
				&$^{+0.5}_{-2.3}$ 
				&$^{+1.9}_{-1.7}$
				&$\pm 4.0$
				&$^{+4.4}_{-4.3}$
				&$\pm 2.7$\\[1ex]  
				\hline  \rule{0pt}{2.5ex}
				{\small \tt NNPDF40an3lo}\cite{NNPDF:2024nan}  
				&$50.6$
				&$-0.1$
				&$^{+0.5}_{-2.3}$ 
				&$^{+0.6}_{-0.6}$
				&$\pm 4.0$
				&$^{+4.0}_{-4.0}$
				&$\pm 1.1$\\[1ex]  
				\hline
			\end{tabular}
		}
		\caption{ Approximate Higgs cross-section at N4LOsv and associated theory errors for $(\muR^c,\muF^c)=(1/2,1/2)\mH$,  $\sqrt{S}=13.6$ TeV, $\alpha_s^c(\MZ) = 0.118$.}
		\label{tab:inc-xsect2} 
	\end{table}
	
\end{itemize}

%%%%%%%%%%%% 
\section{Conclusion}
%%%%%%%%%%%%
We present predictions for the Higgs boson cross-section in gluon-gluon fusion at fourth order in perturbative QCD, incorporating the dominant contributions from soft gluon emissions in the threshold region. 
Soft gluons play a critical role in shaping the perturbative behavior of the cross-section, particularly near threshold. 
At this order, the impact of soft gluon corrections is modest, resulting in a change of approximately $-0.1\%$ relative to N3LO results. 
This small correction indicates excellent perturbative convergence at N4LOsv. 
Furthermore, the theoretical uncertainties associated with a variation of the
renormalization and factorization scales in a wide range are reduced by nearly a factor of two compared to N3LO, yielding highly competitive theoretical predictions suitable for precision comparison with experimental data.

We also assess the residual theoretical uncertainties.
The scale uncertainty, primarily due to soft gluon emission, is reduced to about $2\%$ at this order, down from roughly $4\%$ at N3LO.
To evaluate PDF-related uncertainties, we examine a number of modern PDF sets, determined at NNLO accuracy. 
We find that the associated uncertainties for any given PDF set are generally below $2\%$,
but the cross-section predictions obtained with all PDF sets differ among each other by up to $7\%$.
The dominant source of uncertainty in this repect arises from the strong coupling $\alpha_s$.
From its variation within the $1\sigma$ range, we find an uncertainty of approximately $4\%$.
In addition, we investigate the effect of using the approximate N3LO PDF sets {\tt MSHT20an3lo\_as118} and {\tt NNPDF40\_an3lo\_as\_01180}.
For both sets, we observe a downward shift in the central values of the cross-sections relative to the predictions with NNLO PDFs. 
However, the cross-sections predictions from both sets differ among each other
by $4\%$ despite identical values of $\alpha_s(m_Z)$ used.

In summary, the Higgs boson cross section in gluon-gluon fusion is known very
accurately in perturbative QCD, but the choices of the value for the strong coupling and the
PDF set remain sizable sources of uncertainty, which requires further study
in the future. 
Nevertheless, obtaining the complete N4LO correction would be highly desirable, as it would improve the accuracy of the soft-virtual approximation at this order.

\section*{Acknowledgements}
% TODO: include author contributions
%\paragraph{Author contributions}
%This is optional. If desired, contributions should be succinctly described in a single short %paragraph, using author initials.

% TODO: include funding information
\paragraph{Funding information}
This research is supported by the Deutsche Forschungsgemeinschaft (DFG, German Research Foundation) under grant  396021762 - TRR 257 (\textit{Particle Physics Phenomenology after Higgs discovery}) and   the ERC Advanced Grant 101095857 \textit{Conformal-EIC}.

%%%%%%%%% END TODO: CONTENTS

%%%%%%%%%% TODO: BIBLIOGRAPHY
% Provide your bibliography here. You have two options:

%%% FIRST OPTION
% Write your entries here directly, following the example below, including:
% Author(s), Title, Journal Ref. with year in parentheses at the end, followed by the DOI number.

% \begin{thebibliography}{99}
% \bibitem{1931_Bethe_ZP_71}
% H. A. Bethe, \textit{Zur Theorie der Metalle. i. Eigenwerte und Eigenfunktionen der linearen Atomkette}, Zeit. f{\"u}r Phys. \textbf{71}, 205 (1931), \doi{10.1007\%2FBF01341708}.

% \bibitem{arXiv:1108.2700}
% P. Ginsparg, \textit{It was twenty years ago today...}, (arXiv preprint) \doi{10.48550/arXiv.1108.2700}.

% \end{thebibliography}

%%% SECOND OPTION
% Use your bibtex library, formatted by the SciPost style file.
%\bibliography{SciPost_Example_BiBTeX_File.bib}
\bibliography{ggFn4lo}

\begin{thebibliography}{10}
\providecommand{\url}[1]{\texttt{#1}}
\providecommand{\urlprefix}{URL }
\expandafter\ifx\csname urlstyle\endcsname\relax
  \providecommand{\doi}[1]{doi:\discretionary{}{}{}#1}\else
  \providecommand{\doi}{doi:\discretionary{}{}{}\begingroup
  \urlstyle{rm}\Url}\fi
\providecommand{\eprint}[2][]{\url{#2}}

\bibitem{Dawson:1990zj}
S.~Dawson,
\newblock \emph{{Radiative corrections to Higgs boson production}},
\newblock Nucl. Phys. \textbf{B359}, 283 (1991),
\newblock \doi{10.1016/0550-3213(91)90061-2}.

\bibitem{Djouadi:1991tka}
A.~Djouadi, M.~Spira and P.~M. Zerwas,
\newblock \emph{{Production of Higgs bosons in proton colliders: QCD
  corrections}},
\newblock Phys. Lett. \textbf{B264}, 440 (1991),
\newblock \doi{10.1016/0370-2693(91)90375-Z}.

\bibitem{Spira:1995rr}
M.~Spira, A.~Djouadi, D.~Graudenz and P.~M. Zerwas,
\newblock \emph{{Higgs boson production at the LHC}},
\newblock Nucl. Phys. \textbf{B453}, 17 (1995),
\newblock \doi{10.1016/0550-3213(95)00379-7},
\newblock \eprint{hep-ph/9504378}.

\bibitem{Harlander:2005rq}
R.~Harlander and P.~Kant,
\newblock \emph{{Higgs production and decay: Analytic results at
  next-to-leading order QCD}},
\newblock JHEP \textbf{12}, 015 (2005),
\newblock \doi{10.1088/1126-6708/2005/12/015},
\newblock \eprint{hep-ph/0509189}.

\bibitem{Harlander:2002wh}
R.~V. Harlander and W.~B. Kilgore,
\newblock \emph{{Next-to-next-to-leading order Higgs production at hadron
  colliders}},
\newblock Phys. Rev. Lett. \textbf{88}, 201801 (2002),
\newblock \doi{10.1103/PhysRevLett.88.201801},
\newblock \eprint{hep-ph/0201206}.

\bibitem{Anastasiou:2002yz}
C.~Anastasiou and K.~Melnikov,
\newblock \emph{{Higgs boson production at hadron colliders in NNLO QCD}},
\newblock Nucl. Phys. \textbf{B646}, 220 (2002),
\newblock \doi{10.1016/S0550-3213(02)00837-4},
\newblock \eprint{hep-ph/0207004}.

\bibitem{Ravindran:2003um}
V.~Ravindran, J.~Smith and W.~L. van Neerven,
\newblock \emph{{NNLO corrections to the total cross-section for Higgs boson
  production in hadron hadron collisions}},
\newblock Nucl. Phys. \textbf{B665}, 325 (2003),
\newblock \doi{10.1016/S0550-3213(03)00457-7},
\newblock \eprint{hep-ph/0302135}.

\bibitem{Anastasiou:2015vya}
C.~Anastasiou, C.~Duhr, F.~Dulat, F.~Herzog and B.~Mistlberger,
\newblock \emph{{Higgs Boson Gluon-Fusion Production in QCD at Three Loops}},
\newblock Phys. Rev. Lett. \textbf{114}, 212001 (2015),
\newblock \doi{10.1103/PhysRevLett.114.212001},
\newblock \eprint{1503.06056}.

\bibitem{Mistlberger:2018etf}
B.~Mistlberger,
\newblock \emph{{Higgs boson production at hadron colliders at N$^{3}$LO in
  QCD}},
\newblock JHEP \textbf{05}, 028 (2018),
\newblock \doi{10.1007/JHEP05(2018)028},
\newblock \eprint{1802.00833}.

\bibitem{Moch:2005ky}
S.~Moch and A.~Vogt,
\newblock \emph{{Higher-order soft corrections to lepton pair and Higgs boson
  production}},
\newblock Phys. Lett. \textbf{B631}, 48 (2005),
\newblock \doi{10.1016/j.physletb.2005.09.061},
\newblock \eprint{hep-ph/0508265}.

\bibitem{deFlorian:2014vta}
D.~de~Florian, J.~Mazzitelli, S.~Moch and A.~Vogt,
\newblock \emph{{Approximate N$^{3}$LO Higgs-boson production cross section
  using physical-kernel constraints}},
\newblock JHEP \textbf{10}, 176 (2014),
\newblock \doi{10.1007/JHEP10(2014)176},
\newblock \eprint{1408.6277}.

\bibitem{Agarwal:2021zft}
B.~Agarwal, A.~von Manteuffel, E.~Panzer and R.~M. Schabinger,
\newblock \emph{{Four-loop collinear anomalous dimensions in QCD and N=4 super
  Yang-Mills}},
\newblock Phys. Lett. B \textbf{820}, 136503 (2021),
\newblock \doi{10.1016/j.physletb.2021.136503},
\newblock \eprint{2102.09725}.

\bibitem{Lee:2022nhh}
R.~N. Lee, A.~von Manteuffel, R.~M. Schabinger, A.~V. Smirnov, V.~A. Smirnov
  and M.~Steinhauser,
\newblock \emph{{Quark and Gluon Form Factors in Four-Loop QCD}},
\newblock Phys. Rev. Lett. \textbf{128}(21), 212002 (2022),
\newblock \doi{10.1103/PhysRevLett.128.212002},
\newblock \eprint{2202.04660}.

\bibitem{Falcioni:2024qpd}
G.~Falcioni, F.~Herzog, S.~Moch, A.~Pelloni and A.~Vogt,
\newblock \emph{{Four-loop splitting functions in QCD {\textendash} the
  gluon-gluon case {\textendash}}},
\newblock Phys. Lett. B \textbf{860}, 139194 (2025),
\newblock \doi{10.1016/j.physletb.2024.139194},
\newblock \eprint{2410.08089}.

\bibitem{Das:2019btv}
G.~Das, S.-O. Moch and A.~Vogt,
\newblock \emph{{Soft corrections to inclusive deep-inelastic scattering at
  four loops and beyond}},
\newblock JHEP \textbf{03}, 116 (2020),
\newblock \doi{10.1007/JHEP03(2020)116},
\newblock \eprint{1912.12920}.

\bibitem{Das:2020adl}
G.~Das, S.~Moch and A.~Vogt,
\newblock \emph{{Approximate four-loop QCD corrections to the Higgs-boson
  production cross section}},
\newblock Phys. Lett. B \textbf{807}, 135546 (2020),
\newblock \doi{10.1016/j.physletb.2020.135546},
\newblock \eprint{2004.00563}.

\bibitem{Kniehl:2025ttz}
B.~A. Kniehl, S.~Moch, V.~N. Velizhanin and A.~Vogt,
\newblock \emph{{Flavour Non-Singlet Splitting Functions at Four Loops in QCD
  -- The Fermionic Contributions}}  (2025),
\newblock \eprint{2505.09381}.

\bibitem{Dulat:2018rbf}
F.~Dulat, A.~Lazopoulos and B.~Mistlberger,
\newblock \emph{{iHixs 2 \textemdash{} Inclusive Higgs cross sections}},
\newblock Comput. Phys. Commun. \textbf{233}, 243 (2018),
\newblock \doi{10.1016/j.cpc.2018.06.025},
\newblock \eprint{1802.00827}.

\bibitem{Buckley:2014ana}
A.~Buckley, J.~Ferrando, S.~Lloyd, K.~Nordstr{\"o}m, B.~Page, M.~R{\"u}fenacht,
  M.~Sch{\"o}nherr and G.~Watt,
\newblock \emph{{LHAPDF6: parton density access in the LHC precision era}},
\newblock Eur. Phys. J. C \textbf{75}, 132 (2015),
\newblock \doi{10.1140/epjc/s10052-015-3318-8},
\newblock \eprint{1412.7420}.

\bibitem{Alekhin:2017kpj}
S.~Alekhin, J.~Bl\"umlein, S.~Moch and R.~Placakyte,
\newblock \emph{{Parton distribution functions, $\alpha_s$, and heavy-quark
  masses for LHC Run II}},
\newblock Phys. Rev. D \textbf{96}(1), 014011 (2017),
\newblock \doi{10.1103/PhysRevD.96.014011},
\newblock \eprint{1701.05838}.

\bibitem{Alekhin:2024bhs}
S.~Alekhin, M.~V. Garzelli, S.~O. Moch and O.~Zenaiev,
\newblock \emph{{NNLO PDFs driven by top-quark data}},
\newblock Eur. Phys. J. C \textbf{85}(2), 162 (2025),
\newblock \doi{10.1140/epjc/s10052-025-13832-8},
\newblock \eprint{2407.00545}.

\bibitem{Hou:2019efy}
T.-J. Hou \emph{et~al.},
\newblock \emph{{New CTEQ global analysis of quantum chromodynamics with
  high-precision data from the LHC}},
\newblock Phys. Rev. D \textbf{103}(1), 014013 (2021),
\newblock \doi{10.1103/PhysRevD.103.014013},
\newblock \eprint{1912.10053}.

\bibitem{Bailey:2020ooq}
S.~Bailey, T.~Cridge, L.~A. Harland-Lang, A.~D. Martin and R.~S. Thorne,
\newblock \emph{{Parton distributions from LHC, HERA, Tevatron and fixed target
  data: MSHT20 PDFs}},
\newblock Eur. Phys. J. C \textbf{81}(4), 341 (2021),
\newblock \doi{10.1140/epjc/s10052-021-09057-0},
\newblock \eprint{2012.04684}.

\bibitem{NNPDF:2021njg}
R.~D. Ball \emph{et~al.},
\newblock \emph{{The path to proton structure at 1\% accuracy}},
\newblock Eur. Phys. J. C \textbf{82}(5), 428 (2022),
\newblock \doi{10.1140/epjc/s10052-022-10328-7},
\newblock \eprint{2109.02653}.

\bibitem{PDF4LHCWorkingGroup:2022cjn}
R.~D. Ball \emph{et~al.},
\newblock \emph{{The PDF4LHC21 combination of global PDF fits for the LHC Run
  III}},
\newblock J. Phys. G \textbf{49}(8), 080501 (2022),
\newblock \doi{10.1088/1361-6471/ac7216},
\newblock \eprint{2203.05506}.

\bibitem{McGowan:2022nag}
J.~McGowan, T.~Cridge, L.~A. Harland-Lang and R.~S. Thorne,
\newblock \emph{{Approximate N$^{3}$LO parton distribution functions with
  theoretical uncertainties: MSHT20aN$^3$LO PDFs}},
\newblock Eur. Phys. J. C \textbf{83}(3), 185 (2023),
\newblock \doi{10.1140/epjc/s10052-023-11236-0},
\newblock [Erratum: Eur.Phys.J.C 83, 302 (2023)],
\newblock \eprint{2207.04739}.

\bibitem{NNPDF:2024nan}
R.~D. Ball \emph{et~al.},
\newblock \emph{{The path to $\hbox {N}^3\hbox {LO}$ parton distributions}},
\newblock Eur. Phys. J. C \textbf{84}(7), 659 (2024),
\newblock \doi{10.1140/epjc/s10052-024-12891-7},
\newblock \eprint{2402.18635}.

\bibitem{Cooper-Sarkar:2024crx}
A.~Cooper-Sarkar, T.~Cridge, F.~Giuli, L.~A. Harland-Lang, F.~Hekhorn,
  J.~Huston, G.~Magni, S.~Moch and R.~S. Thorne,
\newblock \emph{{A Benchmarking of QCD Evolution at Approximate $N^3LO$}}
  (2024),
\newblock \eprint{2406.16188}.

\end{thebibliography}

%%%%%%%%%% END TODO: BIBLIOGRAPHY

\end{document}